\documentclass[runningheads]{svmult}

\usepackage{makeidx}   
\usepackage{graphicx}  
\usepackage{subeqnar}  
\usepackage{multicol}  
\usepackage{physprbb}  
\makeindex             

\begin{document}
\title*{The near-infrared view of galaxy evolution}
\toctitle{The Formation and Evolution of Galaxies as seen in
Near-Infrared Surveys}
%
%
\titlerunning{Galaxy Evolution in Near-IR Surveys}
%
\author{Andrea Cimatti}
\authorrunning{Andrea Cimatti}
%
%
\institute{INAF - Osservatorio Astrofisico di Arcetri, Largo E. Fermi 5,
I-50125, Firenze, Italy}

\maketitle              

\begin{abstract}
Near-infrared surveys provide one of the best opportunities to 
investigate the cosmic evolution of galaxies and their mass assembly. 
We briefly review the main results obtained so far with the K20 and 
other recent near-IR surveys on the redshift distribution, the evolution 
of the luminosity function and luminosity density, the nature of old 
and dusty EROs, the evolution of the galaxy stellar mass function, 
the properties of the galaxies in the ``redshift desert'' and the
nature of luminous starbursts at $z\sim2$.
\end{abstract}

\section{Introduction}

Despite the detection of objects up to $z\sim 6.5$ and the impressive
success of the $\Lambda$CDM scenario to account for the properties of
the cosmic microwave background, one of the main and still controversial
questions remains how and when galaxies assembled their mass as a
function of the cosmic time. The hierarchical scenario predicts that 
galaxies built up their present-day mass through a progressive assembly of
smaller sub-systems driven by the merging of dark matter halos. 

With ground-based observations, one of the most solid approaches to address
the problems of galaxy formation and evolution is to study samples of
field faint galaxies selected in the near-infrared, particularly in the
$K$-band (2.2$\mu$m) \cite{bro92,kc98}. Firstly, since the rest-frame 
near-IR luminosity is a good tracer of the galaxy stellar mass \cite{gav96}, 
$K$-band surveys allow to select galaxies according to their mass up 
to $z\sim1.5$ ($\lambda_{rest}\sim 0.9-1.0\mu$m). At higher redshifts, 
the $K$-band starts to sample the rest-frame optical and UV regions, 
and space-based observations at $\lambda_{obs}>2\mu$m are needed to 
cover the rest-frame near-infrared (e.g. SIRTF, ASTRO-F).
Secondly, the similarity of the spectral shapes of different galaxy types 
in the rest-frame near-IR makes the $K$-band selection free from 
strong biases against or in favor of particular classes of galaxies. 
In contrast, the selection of high-$z$ galaxies in the observed optical 
bands is more sensitive to the star formation activity than 
to the stellar mass because it samples the rest-frame UV light and makes 
optical samples biased against old passive or weakly star-forming galaxies. 
Last but not least, near-infrared surveys are less affected by dust extinction 
than optical surveys.

\begin{figure}
\begin{center}
\includegraphics[width=1.\textwidth]{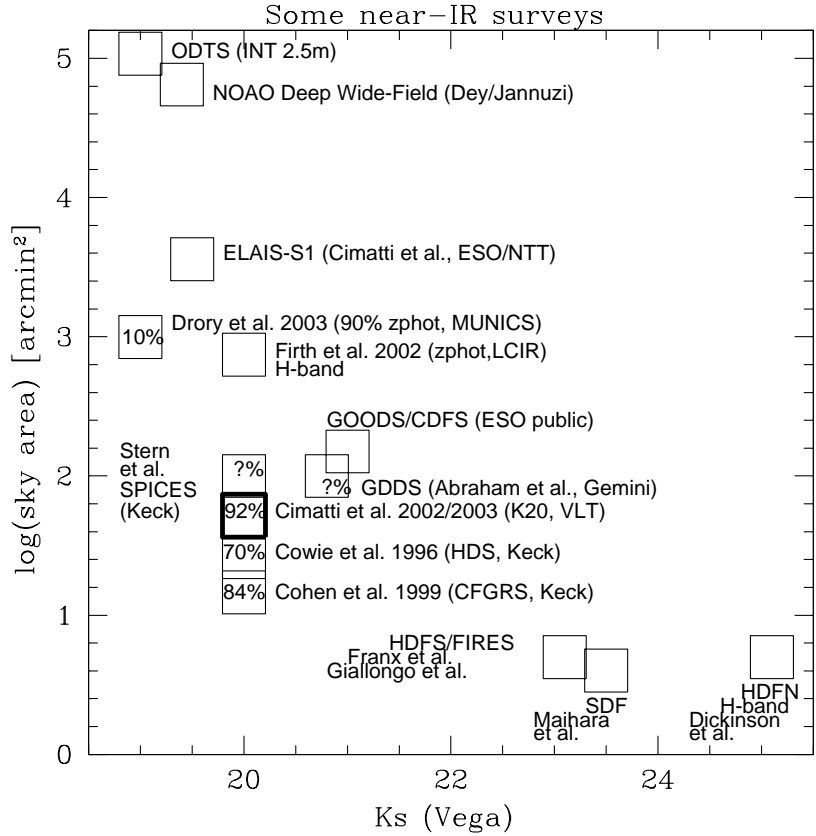}
\end{center}
\caption[]{Some imaging and spectroscopic near-infrared surveys. 
The redshift completeness is shown for spectroscopic surveys.}
\end{figure}

Motivated by the above considerations, several near-infrared surveys
have been undertaken during the last decade \cite{cow96,coh99,cim02a,fir02,lab03,dro03,dic03,fon03} (Fig. 1). 
Spectroscopic surveys are particularly relevant thanks to their 
capability not only to derive the redshifts, but also to unveil the 
nature and the spectral properties of the targeted galaxies. 
Since near-IR multi-object spectrographs are not fully available on 
8-10m-class telescopes, most spectroscopy of $K$-selected galaxies has 
been done in the optical. The most crucial probes of massive 
galaxy evolution are galaxies with the very red colors expected in 
case of passive evolution at $z>1$ (e.g. $R-K_s>5$, usually
called Extremely Red Objects, EROs). However, for the typical 
limiting fluxes of near-IR surveys ($K_s<20-21$), they have very faint optical 
magnitudes ($R\sim25-26$) already close to the spectroscopic limits of 
8-10m-class telescopes. In addition, for $z>1$ their main spectral 
features (e.g. the 4000~\AA~ break and H\&K absorptions) fall 
in the very red part of the observed spectra, where 
the strong OH sky lines and the CCD fringing and low
quantum efficiency make spectroscopy even more demanding. 

\section{The K20 survey}

The K20 survey is a project aimed at investigating galaxy evolution 
through deep ESO VLT spectroscopy of a sample of 546 objects
with $K_s<20$ selected from two fields covering a total area of 52
arcmin$^2$. The two fields are completely independent, and one of them is
a sub-area of the Chandra Deep Field South. Most spectroscopy was done
in the optical with FORS1 and FORS2 and designed to reach the highest
possible signal-to-noise ratio in the red. ISAAC near-IR spectroscopy
was also done for a small fraction of the sample. The multi-band ($UBVRIzJK_s$) 
imaging available for both fields was used to derive high-quality
photometric redshifts ``trained'' with the spectroscopic redshifts.
The final sample covers a redshift range of $0<z<2.5$, and the 
current redshift completeness is 92\% (spectroscopic)
and 100\% (spectroscopic + photometric redshifts). 

The K20 survey represents a significant improvement with respect to
other surveys for faint $K$-selected galaxies thanks to its high 
spectroscopic redshift completeness (the highest to date) extended to
very faint red objects, the larger sample, the coverage of two fields 
and the availability of optimized photometric redshifts. We note here that 
the distribution of the sample over two independent sky fields is 
a significant advantage in reducing the field-to-field variation 
effects. 
For more details on the K20 survey see \cite{cim02a} and 
{\it http://www.arcetri.astro.it/$\sim$k20/}.

\section{Established results: galaxy evolution to $z\sim1$}

Overall, the K20 and other recent near-IR surveys show that galaxies 
are characterized by little evolution to $z\sim1$, so that the observed 
properties can be mimicked by a pure luminosity evolution (PLE)-like 
scenario. This is in contrast with the current $\Lambda$CDM hierarchical 
merging models where the assembly of galaxies occurs later than what 
is actually observed. In particular:

$\bullet$ For $K_s<20$, $N(z)$ has a median redshift of $z_{med}\sim0.8$ 
and a high-$z$ tail extended beyond $z\sim2$. Current hierarchical models 
do not match the observed redshift distribution \cite{cim02c,fir02} (Fig.2), 
but an improved treatment of the merging processes is providing encouraging 
results \cite{som04,men04}.

\begin{figure}
\begin{center}
\includegraphics[width=0.95\textwidth]{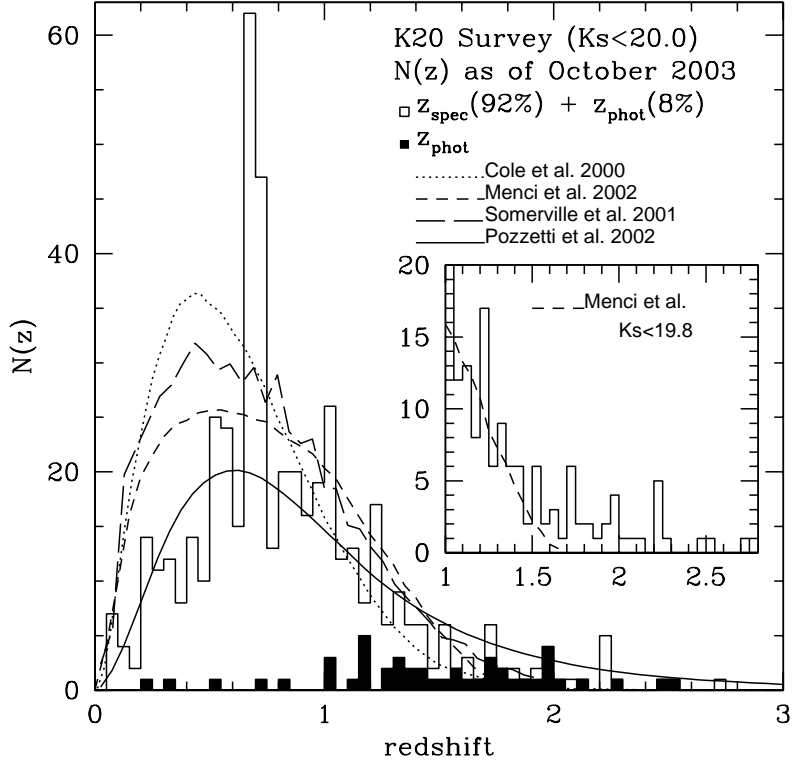}
\end{center}
\caption[]{The K20 survey $N(z)$ compared with 
the predictions of three hierarchical merging model (dotted and 
dashed lines) and a pure luminosity evolution (PLE) model 
(solid line). The models refer to {\it total} magnitudes $K_s<20$ 
and are not corrected for the average under-estimate of 0.2 
magnitudes due to the photometric selection effects evaluated  
in the K20 sample (see \cite{cim02a}).  
Including this bias in the models results in improving  
the agreement with the PLE models and increasing the discrepancy with the
hierarchical models, as shown in the inserted plot where the model
of Menci et al. (2002) for $K_s<19.8$ is compared to the observed
$z>1$ tail of $N(z)$ (see \cite{cim02c}). 
}
\end{figure}

$\bullet$ The rest-frame $K_s$-band luminosity function shows a mild luminosity
evolution up to $z\sim 1$, with a brightening of about 0.5 
magnitudes. Significant density evolution is ruled out up to $z\sim 1$
\cite{coh02,poz03,dro03}. The high-luminosity tail at $z\sim1$ is 
dominated by red early-type galaxies \cite{poz03}.
Current hierarchical models fail in reproducing the shape
and the evolution of the LF as they over-predict the number of
sub-$L^{\ast}$ galaxies, under-predict that of luminous galaxies,
and predict a strong density evolution.

$\bullet$ The rest-frame $K_s$-band luminosity density evolves slowly to
$z\sim1$ with $\rho_{K_s}(z) \propto (1+z)^{0.37}$, and much slower 
than the UV luminosity density \cite{poz03}. 

$\bullet$ The properties of ``old'' EROs with $K_s<20$ 
at $z\sim1$ (morphology, spectra, 
luminosities, ages, stellar masses, clustering) imply the existence of a 
substantial population of old (a few Gyr), passively evolving and fully 
assembled massive spheroids which requires that major episodes of 
massive galaxy formation occurred at least at $z_{form}\sim2$ 
\cite{cim02b,cim03}. Their number density at $z\sim1$ is 
consistent (within $2\sigma$)
with that of local luminous E/S0 galaxies. Current hierarchical models 
severely under-predict the number of ``old'' EROs \cite{cim02b,cim03}.

$\bullet$ A numerous population of dusty star-forming EROs with disk-like and
irregular morphologies emerges at $0.7<z<1.7$ from near-IR surveys
\cite{cim02b,sma02,cim03}. These objects are often too faint to be 
detected in submm surveys due to their inferred far-IR luminosities 
$<10^{12}$ L$_{\odot}$ and represent an ensemble of galaxies important
to complement other star-forming systems selected with different
techiniques. They are also expected to be important contributors to the 
cosmic star-formation density \cite{cim02b,cim03,bru02,sma02}. Also in 
this case, current hierarchical models under-estimate the number of ``dusty''
EROs.

$\bullet$ ``Old'' EROs seem to have much stronger clustering than ``dusty''
EROs, with a comoving $r_0$ similar to that of present-day luminous ellipticals
\cite{dad02}.

$\bullet$ The number density of massive galaxies and the cosmic stellar mass
density $\Omega_{\ast}$ show a slow decrease from $z\sim0$ to $z\sim1$
qualitatively consistent with a hierarchical scenario, but much slower than
what the current models predict \cite{fon04,bri00,dro01,coh02,dic03,fon03}.

\section{Beyond $z\sim1$: galaxies in the ``redshift desert''}

At higher redshifts, the picture becomes more controversial:
at $z>1$ the evolution of the near-IR luminosity function seems 
to depart from a PLE-like pattern \cite{poz03}, it is still not clear
if the number density of old spheroids drops at $z>1.3$ 
\cite{ben99,rod01}, and unexpected populations of massive 
star-forming galaxies are being found in the range of $1.5<z<3$ 
thanks to near-IR surveys \cite{fra03,van03,dad04}.

Moreover, the galaxy stellar mass function and $\Omega_{\ast}(z)$ 
display a fast evolution at $z>1$ \cite{dic03,fon03}, and the K20 
survey results suggest a rapid increase of the stellar mass density 
from $20^{+20}_{-10}$\% of the local value at $2<z<3$ to about 100\%
at $z\leq1$ \cite{fon04}. 

These results suggest that $1.5<z<2.5$ may be the critical cosmic epoch 
during which most star formation activity and galaxy mass assembly 
took place. However, probing the $1.5<z<2.5$ range is difficult due 
to the lack of strong spectral features and emission lines redshifted 
in the observed optical and to the strong OH emission sky lines severely 
affecting the spectra beyond $\sim$8000~\AA. Due to the limited number of 
spectroscopically identified galaxies, this redshift range has been
traditionally called ``redshift desert''.

Thanks to the very deep red-optimized optical spectroscopy and to  
near-IR spectroscopy, the K20 survey allowed us to populate the
``redshift desert'' and to unveil the nature of $K$-selected
galaxies in this critical redshift range. By combining deep 
HST+ACS imaging with
K20 spectroscopy, we found that a variety of galaxy types are present
at $1.5<z<2.5$: spheroids at $z\sim1.5$, disk- and
spiral-like systems, and many irregular merging-like objects.

Particularly relevant is a sub-sample of luminous galaxies  
at $1.7<z<2.3$ spectroscopically identified in the K20 survey
\cite{dad04}. Their inferred star formation rates 
of $\sim 100-500$ M$_{\odot}$yr$^{-1}$, stellar masses up to $10^{11}$
M$_{\odot}$ derived from the fitting of their multi-color spectral
energy distributions, and their merging-like optical morphologies
becoming more compact in the near-IR, suggest that these galaxies
may be massive galaxies caught during their major episode of
mass assembly and star-formation, possibly being the progenitors
of the present-day massive spheroidal systems. Current semi-analytical 
models under-predict by a factor of $\sim30$ the 
number density of such galaxies \cite{dad04,som04}.
Other near-IR surveys are independently finding candidate massive 
systems at $z\sim2-3$ \cite{fra03,van03}.

In order to draw a coherent picture of galaxy evolution at $1.5<z<2.5$, 
it will be crucial to combine $K$-selected samples (e.g. K20, FIRES
\cite{fra03}, GDDS \cite{gla03}) with UV- and optically-selected 
samples more biased toward 
star-forming objects with little dust extinction such as those resulting 
from the Lyman-break selection at $z>1.4$ \cite{ste}.

\section{Acknowledgments}
The K20 survey team includes: T. Broadhurst (HUJ),
S. Cristiani (INAF-Trieste), S. D'Odorico
(ESO), E. Daddi (ESO), A. Fontana (INAF-Roma), E. Giallongo (INAF-Roma), 
R. Gilmozzi (ESO), N.  Menci (INAF-Roma), M. Mignoli (INAF-Bologna), 
F. Poli (University of Rome), L. Pozzetti (INAF-Bologna), A. Renzini 
(ESO), P. Saracco (INAF-Brera), and G. Zamorani (INAF-Bologna).

%

\end{document}